\documentclass[arxiv,fleqn]{w-art-mod}
\usepackage{times}
\usepackage{w-thm}
\usepackage[authoryear]{natbib}
\theoremstyle{plain}

\theoremstyle{definition}

\usepackage[]{graphicx}
\chardef\bslash=`\\ 

\usepackage{amsfonts}
\usepackage{graphicx}
\usepackage[colorlinks=true, allcolors=black]{hyperref}

\usepackage{lineno}
\usepackage[utf8]{inputenc}
\usepackage{color, colortbl}
\definecolor{gray}{gray}{0.7}
\definecolor{lightgray}{gray}{0.9}
\usepackage{hyperref}
\usepackage{xcolor}
\usepackage{natbib}

\begin{document}
\DOIsuffix{}
\Volume{}
\Issue{}
\Year{2023}
\pagespan{1}{}
\keywords{Conservative tests;  discretely distributed test statistics; group testing; multiple comparisons; randomized tests.\\
\noindent \hspace*{-4pc} 
}  

\title[Randomized $p$-values in discrete models]{Multiple testing of composite null hypotheses for discrete data using randomized $p$-values}

\author[OCHIENG ET AL.]{Daniel Ochieng} 
\address[\inst{1}]{Institute for Statistics, University of Bremen, 28344 Bremen, Germany.}
\author[]{Anh-Tuan Hoang} 
\author[]{Thorsten Dickhaus 
\footnote{Corresponding author: {\sf{e-mail: dickhaus@uni-bremen.de}}}}
\Receiveddate{} \Reviseddate{} \Accepteddate{}

\begin{abstract}


$P$-values that are derived from continuously distributed  test statistics are typically uniformly distributed on $(0,1)$ under least favorable parameter configurations (LFCs) in the null hypothesis. Conservativeness of a $p$-value $P$ (meaning that $P$ is under the null hypothesis stochastically larger than a random variable which is uniformly distributed on $(0,1)$) can occur if the test statistic from which $P$ is derived is discrete, or if  the true parameter value under the null is not an LFC. To deal with both of these sources of conservativeness, we present two approaches utilizing randomized $p$-values, namely single-stage and two-stage randomization. We illustrate their effectiveness for testing a composite null hypothesis under a binomial model. We also give an example of how the proposed $p$-values can be used to test a composite null in group testing designs. Similar to previous findings, we find that the proposed randomized $p$-values are less conservative compared to non-randomized $p$-values under the null hypothesis, but that they are stochastically not smaller under the alternative. The problem of establishing the validity of randomized $p$-values is not trivial and has received attention in previous literature. We show that our proposed randomized $p$-values are valid under various discrete statistical models which are such that the distribution of the corresponding test statistic belongs to an exponential family. The behaviour of the power function for the tests based on the proposed randomized $p$-values as a function of the sample size is also investigated. Simulations and a real data analysis are used to compare the different considered $p$-values.

\end{abstract}

\maketitle                   







\section{INTRODUCTION}
\label{s:intro}

In modern statistical data analysis, for example in genomics (cf.\ \cite{dudoit2008multiple}, ch. 9-12), genetics and other life sciences (cf.\ \cite{dickhaus2014simultaneous}, ch. 9-12), or finance (cf.\ \cite{harvey2020evaluation}), the practitioner is often interested in testing thousands of null hypotheses simultaneously, which requires a multiplicity adjustment. Furthermore, some of these applications, like genetic association studies or clinical trials with categorical outcomes, generate discrete data in terms of counts, which leads to test statistics which follow a discrete distribution. Consequently, $p$-values which are derived as deterministic transformations of such test statistics are also discrete, and thus non-uniform under the null.
Some suggestions for dealing with discretely distributed test statistics in the multiple testing context have been made in previous literature. They include, among others, (i) utilization of mid $p$-values or randomized $p$-values (see, e.\ g., \cite{tocher1950extension, habiger2011randomised}),  
and (ii) data-adaptive procedures based on estimators of the proportion of true null hypotheses, where the estimator is tailored to the discrete model (see, e.\ g., \cite{chen2018multiple}).
The assumption of uniformity of the $p$-value under the null also fails to hold true in the case of a composite null hypothesis (containing more than one parameter value) if the true parameter value is not a least favorable configuration (LFC). Typically, LFCs are located at the boundary of the null hypothesis, thus the $p$-value is non-uniform especially if the true parameter value lies "deep inside" the null. Randomization has been proposed as a possible solution by \cite{dickhaus2013randomized,hoang2021usage,hoang2022randomized}. In this work, we focus on randomization to address both of the aforementioned reasons for non-uniform $p$-values under the null. 

\subsection{Related work}
The idea of randomization was first considered by \cite{tocher1950extension} who suggested it as a corrective measure after showing that a discrete test may not exactly achieve a given significance level since the resulting $p$-value
is not continuously distributed.  \cite{lancaster1961significance} proposed the mid  $p$-value by replacing the uniform random variable involved in randomization by its expected value. 
%
%
\cite{geyer2005fuzzy} introduced fuzzy or abstract randomized $p$-value and fuzzy confidence intervals. 
They also noted the ``unavoidable flaw'' that the coverage probability as a function of the parameter value exhibits an oscillatory behavior for any non-randomized confidence interval (or crisp interval) for discrete data, as depicted in Figure $1$ of \cite{geyer2005fuzzy}. 
This oscillation of the coverage probability and of the power function for test statistics with a discrete distribution also occurs when regarded as a function of the sample size (cf. \cite{finner2001increasing,finner2001ump,thulin2014split}), which appears paradoxical and is very inconvenient in practice.

\cite{dickhaus2012analyze} considered exact association tests in contingency tables (conditional to the marginal counts). Non-randomized versions of these tests are reproducible, but they fail to exhaust the significance level, which typically also implies non-optimal power. The authors proposed the utilization of randomized $p$-values, especially for estimating the proportion of the true null hypotheses in cases where many tables have to be analyzed simultaneously. 
Multiplicity-adjusted versions of the randomized $p$-value,  the mid $p$-value, and the abstract randomized $p$-value  were proposed by \cite{habiger2015multiple} for test statistics that have a discrete distribution. These discretely distributed test statistics can arise, for example, when using non-parametric rank-based methods. 
In the case of multiple testing of heterogeneous discrete data, \cite{dai2019non} proposed a procedure based on the marginal critical function (MCF) of randomized tests. The MCF approach provides a non-random decision that is based on ranked MCF values, although it utilizes randomized tests. 

Different randomization techniques have also been proposed for the construction of confidence intervals.  A randomization technique called the split sample method was proposed by \cite{decrouez2014split} and \cite{thulin2014split} for discrete data. The sample is randomly divided into two parts and the maximum likelihood estimate (MLE) involved in the confidence interval construction is replaced by a weighted estimate that results from the two sub-samples. Splitting the sample into two parts helps in reducing the oscillation in the coverage probability of the confidence interval for a binomial parameter. 
\cite{korn1987data} considered a procedure called “data-randomization”, which utilizes the data themselves to generate the randomization. An extension of the \cite{stevens1950fiducial} interval and additional features of the \cite{korn1987data} interval were considered by \cite{kabaila2013randomized}, who also discussed objections to the concept of randomization and some potential corrective measures. An exact randomized interval based on a prior-free probabilistic inference model was proposed by \cite{lu2019prior}. They found that the interval has exact coverage probability, avoiding the oscillatory behavior shown by non-randomized intervals.

Other research that has utilized the general concept of randomization includes \cite{dunn1996randomized} who used randomization to obtain continuous residuals for a discrete response variable in regression analysis. Normalized randomized survival probabilities for model diagnostics in regression were proposed by \cite{li2021model}. 

\subsection{Main contributions}
In this work, we extend the (single-stage) randomization procedures of \cite{hoang2021usage} and \cite{hoang2022randomized} to the case of discrete models. This is done by introducing a minor modification of the procedure. Furthermore, we propose the following novel two-stage randomization approach: We first take care of the discreteness of the $p$-values by applying the randomization proposed by \cite{dickhaus2012analyze}. Under LFCs, this leads to exactly uniformly distributed randomized $p$-values. However, these $p$-values are still conservative under non-LFCs. Therefore, we apply the approach  by \cite{hoang2021usage,hoang2022randomized} on these $p$-values in the second stage. 
We show that for a discretely distributed test statistic, two-stage randomization results in a $p$-value that is in distribution closer to uniformity under the null hypothesis as compared to  single-stage randomization. In particular, this leads to a more accurate estimation of the proportion of true null hypotheses in multiple testing. 
Furthermore, we prove the validity of the proposed randomized $p$-values for test statistics which are likelihood ratio ordered, for example, in a one-parametric exponential family of distributions. For the two-stage randomized $p$-values to be valid, it is required that their non-randomized versions are so-called uniformly valid, cf. \cite{Whitt1980,Whitt1982}, \cite{Lynch1987}, and \cite{zhao2019multiple}, among others.

We exemplify the usage of the proposed randomized $p$-values with testing for the success parameter under a binomial model. 
%
When testing composite null hypotheses for the binomial success parameter, \cite{finner2001ump} showed that the power function of the uniformly most powerful (UMP) test (at a fixed significance level) does not monotonically increase with the sample size $n$. We demonstrate that this issue can be addressed by randomization. A power function which is monotonically increasing in $n$ facilitates sample size planning, because it is guaranteed that additional observational units cannot lower the power.

Finally, we present computer simulations and a real data analysis to evaluate the performance of the proposed $p$-values in a multiple testing context, where the proportion of true null hypotheses is estimated with the method of \cite{schweder1982plots}.
\subsection{Overview of the rest of the material}

The rest of this paper is organized as follows. General preliminaries are provided in Section $\ref{general-assum}$. In Section $\ref{s:model}$, we describe the proposed randomization procedures for a general discrete model and a composite null hypothesis. In Section $\ref{s:binomial}$, we calculate the one- and two-stage randomized $p$-values, as well as their cumulative distribution functions (CDFs), for the general binomial model. In the same section,  we compare the power of the test based on the randomized $p$-values for different sample sizes and give an example in group testing when there is the possibility of group misclassification. Estimation of the proportion of true null hypotheses in multiple testing is considered in Section $\ref{s:multiple}$. Finally, we discuss our results in Section $\ref{s:discussion}$.


\section{GENERAL PRELIMINARIES}\label{general-assum}

We assume that our (random) data is given by $\pmb{X}=(X_1,\ldots,X_n)^\top$, where each $X_i$ is a real-valued, observable random variable, $1 \leq i \leq n$, and all $X_i$ are stochastically independent and identically distributed (i.i.d.). The support of $\pmb{X}$ will be denoted by $\mathcal{X}$. We assume that the marginal distribution of $X_1$ is given by $P_\theta$, where $\theta \in \Theta \subseteq \mathbb{R}$ is the parameter of the statistical model. The distribution of $\pmb{X}$ under $\theta$ is consequently given by $P_\theta^{\otimes n} =: \mathbb{P}_\theta$. We will be concerned with one-sided test problems of the form
\begin{equation}
H: \theta\leq \theta^* \text{~~versus~~} K: \theta> \theta^*, \label{testproblem}
\end{equation} 
where $\theta^*$ is a pre-specified constant.
We consider test statistics $T(\pmb{X})$, where $T: \mathcal{X} \to \mathbb{R}$ is a measurable mapping. We assume that the marginal $p$-value $p(\pmb{X})$ derived from $T(\pmb{X})$ 
is valid, meaning that $\mathbb{P}_\theta(p(\pmb{X}) \leq \alpha)\leq \alpha$ holds true under any parameter value $\theta$ in the  null hypothesis and for all $\alpha\in [0,1]$. If in this condition a parameter value $\theta_{LFC}$ maximizes the left-hand side for all $\alpha$, we call $\theta_{LFC}$ an LFC. Valid $p$-values are under the null stochastically not smaller than the uniform distribution on the unit interval $[0, 1]$, which we denote by UNI$[0,1]$. Especially in the case of discrete models, valid $p$-values are typically strictly stochastically larger than UNI $[0,1]$, as investigated by, among many others,  \cite{finner2007note}, \cite{habiger2011randomised}, \cite{dickhaus2012analyze}, and \cite{habiger2015multiple}. 

According to \cite{zhao2019multiple}, a $p$-value $p(\pmb{X})$ is called uniformly valid for testing $H$, if for all $\theta \in H$ and for all $\tau \in (0, 1]$ fulfilling  $\mathbb{P}_\theta(p(\pmb{X}) \leq \tau) > 0$, it holds that
\begin{equation}\label{uniformly-valid}
\forall t \in [0, \tau]: \frac{\mathbb{P}_\theta(p(\pmb{X}) \leq t)}{\mathbb{P}_\theta(p(\pmb{X}) \leq \tau)} \leq \frac{t}{\tau}.
\end{equation}
It is known that uniform validity in the sense of \eqref{uniformly-valid} holds for $p$-values corresponding to one-sided tests for problems of the form \eqref{testproblem}, if the distributions of the test statistic have monotone likelihood ratios with respect to $\theta$, cf. \cite{zhao2019multiple}. For example, one-sided tests in the binomial model that we will be investigating in Section \ref{s:binomial} have this property, due to the structure of a one-parametric exponential family for the likelihood functions. We will provide further details on this in Section \ref{s:properties}. Of course, uniform validity in the sense of \eqref{uniformly-valid} implies validity by virtue of considering $\tau=1$ in \eqref{uniformly-valid}.

For the theoretical analysis in Section \ref{s:properties}, we will make use of concepts of stochastic orders as treated, e.\ g., in the monograph by \cite{shaked2007stochastic}. In particular, we will use the symbol $\leq_{st}$ to denote the (usual) stochastic order, and the symbol $\leq_{rh}$ to denote the reverse hazard rate order. It is known that the likelihood ratio order is stronger than the reverse hazard rate order; see Theorem 1.C.1 in \cite{shaked2007stochastic}.

Finally, we will need in the sequel the (generalized) inverses of certain non-decreasing functions mapping from $\mathbb{R}$ to $[0, 1]$. To this end, we follow Appendix 1 in \cite{reiss1989}: If $F$ is a real-valued, non-decreasing, right-continuous function defined on $\mathbb{R}$, we let $F^{-1}(y) = \inf\{x \in \mathbb{R}: F(x) \geq y\}$. If $G$ is a real-valued, non-decreasing, left-continuous function defined on $\mathbb{R}$, we let $G^{-1}(y) = \sup\{x \in \mathbb{R}: G(x) \leq y\}$.


\section{RANDOMIZATION PROCEDURES}
\label{s:model}
\subsection{Introduction}
As mentioned before, we will consider both single- and two-stage randomization procedures. For single-stage randomization, a modification of the randomized $p$-value given by \cite{hoang2021usage,hoang2022randomized} is utilized. For the two-stage randomization, the first stage is carried out using the approach proposed in Appendix II of \cite{dickhaus2012analyze}. This transforms the discrete $p$-value based on $T$ into a continuous $p$-value. In the second stage, the approach given by \cite{hoang2021usage} is applied to deal with the conservativeness of the $p$-value resulting from the first stage. This conservativeness is  due to the composite nature of the hypothesis $H$. We first give a description of the single-stage procedure and then one for two-stage randomization.

\subsection{Single-stage randomization}
Suppose we wish to randomize only once. In that case, we can make use of the single-stage randomized $p$-value defined by \cite{hoang2021usage} with a slight modification. The LFC-based $p$-value for our hypothesis is defined as
\begin{equation}
 \hspace{100pt minus 1fil} P^{LFC}(\pmb{X})=1-F_{\theta^*}(T(\pmb{X})-),
\label{eq:lfc1}
\end{equation}
where  $F_{\theta^*}$ is the CDF of the test statistic $T$ under the LFC  $\theta^*$ and $f(x-):= \lim_{y\uparrow x} f(y)$ for a function $f:\mathbb{R}\to \mathbb{R}$. We use $p$-LFC and $P^{LFC}(\pmb{X})$ interchangeably to denote the $p$-value defined in
$(\ref{eq:lfc1})$. The CDF of $p$-LFC is given by
\begin{equation}
  \hspace{90pt minus 1fil}   \mathbb{P}_{\theta}\left(P^{LFC}(\pmb{X})\leq t\right)=1-F_{\theta}(G^{-1}_{\theta^*}(1-t)-),
    \label{eq:lfc3}
\end{equation}
where $G_{\theta^*}(x) = F_{\theta^*}(x-)$, $x \in \mathbb{R}$.
\cite{hoang2021usage} defined a randomized $p$-value as follows: Let $U$ be a UNI$[0,1]$-distributed random variable that is stochastically independent of $\pmb{X}$. For a given constant $c\in (0,1]$, the randomized $p$-value $P^{rand1}(\pmb{X},U,c)$ based on $P^{LFC}(\pmb{X})$ is then defined as
\begin{equation}
\hspace{20pt minus 1fil} P^{rand1}(\pmb{X},U,c)= U \pmb{1}\{P^{LFC}(\pmb{X})\geq c\}
+ \frac{P^{LFC}(\pmb{X})}{c^*}\pmb{1}\{P^{LFC}(\pmb{X})<c\},
\label{eq:r4}
\end{equation}
and $P^{rand1}(\pmb{X},U,0)=U$. We propose to let  $c^*=\mathbb{P}_{\theta^*}\{P^{LFC}(\pmb{X})<c\}$ (instead of $c^*=c$ as in \cite{hoang2021usage}), which is our proposed modification for discretely distributed $p$-value. If $P^{LFC}(\pmb{X})$ is UNI$[0,1]$-distributed under the LFC parameter, then both versions are equivalent. Notice that $c^*$ is the largest support point
of $P^{LFC}(\pmb{X})$ under an LFC parameter $\theta^*$ such that $c^*\leq c$. The interpretation of $P^{rand1}(\pmb{X},U,c)$ is such that, when $P^{LFC}(\pmb{X})<c$ then $P^{LFC}(\pmb{X})$ is transformed by dividing it by $c^*$, otherwise it is replaced by the random variable $U$. The division by $c^*$ is needed to ensure (conditional) validity of $P^{rand1}(\pmb{X},U,c)$, and it has its conceptual origins in the general approach to selective inference; see the discussion around Definition $3.1$ in \cite{hoang2022randomized}. The CDF of $P^{rand1}(\pmb{X},U,c)$ is given by
\begin{equation}
\begin{split}
\hspace{20pt minus 1fil} \mathbb{P}_{\theta}\{P^{rand1}(\pmb{X},U,c)\leq t\} &= t\mathbb{P}_{\theta}\{P^{LFC}(\pmb{X})>c\}
+\mathbb{P}_{\theta}\{P^{LFC}(\pmb{X})\leq tc^*\}\label{eq:r5},\\
&=\mathbb{P}_{\theta}\{P^{rand1}(\pmb{X},U,c^*)\leq t\}.
\end{split}
\end{equation}
The second equality in $(\ref{eq:r5})$ holds due to $c^*\leq c$, so that $\mathbb{P}_{\theta}\{P^{LFC}(\pmb{X})>c\}=\mathbb{P}_{\theta}\{P^{LFC}(\pmb{X})>c^*\}$ (assuming the support points of $p$-LFC do not depend on $\theta$), since there is no other support point between $c$ and $c^*$. Single-stage randomization deals with the discreteness of the model and the composite nature of the hypothesis all at once. However, it fails to remove the conservativity of $P^{LFC}(\pmb{X})$ arising from the discreteness of the model completely. This will be seen more clearly in the plots in Section $\ref{s:binomial}$. This is the motivation for the two-stage randomization procedure, which we describe next.

\subsection{Two-stage randomization}
We now turn our attention to the two-stage randomization procedure and describe it in detail. For the first stage, we make use of the following randomized $p$-value for discrete models from \cite{dickhaus2012analyze}: Assume that the real-valued test statistic $T$ tends to larger values under the alternative. Assume $U$ to be a UNI$[0,1]$-distributed random variable which is stochastically independent of $\pmb{X}$. Then, the randomized $p$-value pertaining to $T$ that we are considering in the sequel is given by
\begin{equation}
\hspace{50pt minus 1fil} P_T^{rand}(\pmb{X},U)=\sum_{\pmb{y}:T(\pmb{y})> T(\pmb{X})}
f_{\theta^*}(\pmb{y})+ U \sum_{\pmb{y}:T(\pmb{\pmb{y}})=T(\pmb{X})}f_{\theta^*}(\pmb{y}),
\label{eq:r1}
\end{equation}
where $f_{\theta^*}$ denotes the probability mass function (pmf) of $\pmb{X}$ under $\theta^*$.
The CDF of $P_T^{rand}$ is given by
\begin{equation}
\hspace{70pt minus 1fil}    \mathbb{P}_{\theta}\{P_T^{rand}(\pmb{X},U)\leq t\}=
1-F_{\theta}(y(t)) + g(t) f_{\theta}(y(t)),
\label{eq:w1}
\end{equation}
where $g(t)=\{F_{\theta^*}(y(t))-(1-t)\}\{f_{\theta^*}(y(t))\}^{-1}$, $y(t)=F^{-1}_{\theta^*}(1-t)$, and where $F_{\theta^*}$ and $F^{-1}_{\theta^*}$ are the CDF and the quantile function of $T$ under the LFC parameter $\theta^*$, respectively. We note that thresholding $P_T^{rand}$ at $\alpha$ for making a test decision is under certain conditions equivalent to utilizing the well-known (potentially randomized) UMP level $\alpha$ test based on $T$ (which is in such cases a deterministic transformation of the likelihood ratio of the statistical model under consideration). In the case of a one-sided test under a binomial model, the aforementioned equivalence holds true. Therefore, we sometimes refer to $P_T^{rand}$ as the UMP $p$-value in our context. 

In discrete models, non-randomized $p$-values are usually conservative, that is, under the null hypothesis they are valid and $\mathbb{P}(p(\pmb{X})\leq \alpha)< \alpha$ for some $\alpha\in (0,1)$. 
Tests conducted using these $p$-values will fail to exhaust the significance level. 
This non-uniformity is not a problem for single tests, but can lower the multiple power of multiple tests, as noted by \cite{dickhaus2012analyze}. Our proposed solution to this is to use the $p$-value $P_T^{rand}$ in a first stage of randomization, in order to transform the discrete test statistic into a continuously distributed $p$-value. 

The proposed second stage of randomization using the randomization technique from \cite{hoang2021usage} can now be applied to deal with the conservativeness of  $P_T^{rand}$ that results from the composite nature of our null hypothesis. To this end, let $\tilde{U}$ be another UNI$ [0,1]$-distributed random variable which is stochastically independent of the data $\pmb{X}$ and stochastically independent of $U$. Assume also that a constant $c\in (0,1]$ is given. The randomized $p$-value $P^{rand2}(\pmb{X},U,\tilde{U},c)$ in the second stage is defined as
\begin{equation}
P^{rand2}(\pmb{X},U,\tilde{U},c)=\tilde{U} \pmb{1}\{P_T^{rand}(\pmb{X},U)\geq c\}+P^{rand}_T(\pmb{X},U)(c)^{-1}\pmb{1} \{P^{rand}_T(\pmb{X},U)<c\}. \label{eq:r2}
\end{equation}
Furthermore, we define $P^{rand2}(\pmb{X},U,\tilde{U},0)=\tilde{U}$. The CDF of $P^{rand2}$ is given by
\begin{equation}
\mathbb{P}_{\theta}\{P^{rand2}(\pmb{X},U,\tilde{U},c)\leq t\}=t\mathbb{P}_{\theta}\{P_T^{rand}(\pmb{X},U)>c\} 
+\mathbb{P}_{\theta}\{P_T^{rand}(\pmb{X},U)\leq tc\} \label{eq:r3}.
\end{equation}

\subsection{Properties of the proposed \texorpdfstring{$p$}{p}-values}\label{s:properties}
Under parameter values in the  null hypothesis, randomized $p$-values are typically in distribution closer to UNI$[0, 1]$ than non-randomized $p$-values (cf. \cite{dickhaus2013randomized}). For easier reference, Table \ref{t:tableone} provides an overview of all the $p$-values described in this section, together with their corresponding CDFs.

\begin{table}[htb]
\begin{center}
  \caption{Summary of the LFC-based $p$-value ($p$-LFC), UMP $p$-value (PT-RAND), single-stage randomized $p$-value (RAND1) based on $p$-LFC, and two-stage randomized $p$-value (RAND2) based on PT-RAND, with their corresponding cumulative distribution functions (CDFs).}
\label{t:tableone}
\begin{tabular}{lc}
\rowcolor{gray}  
& $p$-LFC \\[0.2cm]
\rowcolor{lightgray}
$p$-value (Eq. $\eqref{eq:lfc1}$)& $P^{LFC}(\pmb{X})=1-F_{\theta^*}(T(\pmb{X})-)$\\[0.3cm]
CDF (Eq. \eqref{eq:lfc3}) &  $\mathbb{P}_{\theta}\left(P^{LFC}(\pmb{X})\leq t\right)=1-F_{\theta}(G^{-1}_{\theta^*}(1-t)-)$,
where $G_{\theta^*}(x) = F_{\theta^*}(x-)$, $x \in \mathbb{R}$.\\[0.3cm]
\rowcolor{gray} 
&RAND1 \\ [0.2cm]
 \rowcolor{lightgray}
 $p$-value (Eq. $\eqref{eq:r4}$)&$P^{rand1}(\pmb{X},U,c)=U \pmb{1}\{P^{LFC}(\pmb{X})\geq c\}+P^{LFC}(\pmb{X})\{c^*\}^{-1}$$\pmb{1}\{P^{LFC}(\pmb{X})<c\}$ \\[0.3cm]
CDF (Eq. $\eqref{eq:r5}$)& $\mathbb{P}_{\theta}\{P^{rand1}(\pmb{X},U,c)\leq t\}=t\mathbb{P}_{\theta}\{P^{LFC}(\pmb{X})>c\}+\mathbb{P}_{\theta}\{P^{LFC}(\pmb{X})\leq tc^*\} $\\ [0.3cm]  
\rowcolor{gray} 
& PT-RAND \\[0.2cm]
\rowcolor{lightgray} 
$p$-value (Eq. $\eqref{eq:r1}$)&$P_T^{rand}(\pmb{X},U)=\sum_{\pmb{y}:T(\pmb{y})> T(\pmb{X})}f_{\theta^*}(\pmb{y})$$+U \sum_{\pmb{y}:T(\pmb{y}) = T(\pmb{X})}f_{\theta^*}(\pmb{y})$\\[0.3cm]
CDF (Eq. \eqref{eq:w1})&$\mathbb{P}_{\theta}\{P_T^{rand}(\pmb{X},U)\leq t\}$=
$1-F_{\theta}(y(t))\nonumber+g(t) f_{\theta}(y(t))$,\\
& $g(t)=\{F_{\theta^*}(y(t))-(1-t)\}\{f_{\theta^*}(y(t))\}^{-1}$,\\
&$y(t)=F^{-1}_{\theta^*}(1-t)$\\[0.3cm]
\rowcolor{gray} 
&RAND2 \\ [0.2cm]
\rowcolor{lightgray} 
$p$-value (Eq. $\eqref{eq:r2}$) &$P^{rand2}(\pmb{X},U,\tilde{U},c)=\tilde{U} \pmb{1}\{P_T^{rand}(\pmb{X},U)\geq c\}+P^{rand}_T(\pmb{X},U)(c)^{-1}\pmb{1} \{P^{rand}_T(\pmb{X},U)<c\}$\\[0.3cm]
CDF (Eq. $\eqref{eq:r3}$) &$\mathbb{P}_{\theta}\{P^{rand2}(\pmb{X},U,\tilde{U},c)\leq t\}$$=t\mathbb{P}_{\theta}\{P_T^{rand}(\pmb{X},U)>c\}+\mathbb{P}_{\theta}\{P_T^{rand}(\pmb{X},U)\leq tc\}$\\[0.3cm] 
\end{tabular}
\end{center}
\end{table}

The remainder of this section is devoted to establishing the validity of the proposed $p$-values. To this end, we first present an auxiliary lemma.

\begin{lemma}\label{s:validityrand1}
Let $\mathcal{M}$ be a subset of $\mathbb{R}$, let $\tau\in(0,1]$ be a given constant, and let $\theta^{\star}$ be a fixed parameter value. Let $P$ be a $p$-value, and assume that the following three conditions are fulfilled: (i) $\mathbb{P}_{\theta^{\star}}\big(P\leq \mathbb{P}_{\theta^{\star}}\left(P\leq\tau\right)\big)=\mathbb{P}_{\theta^{\star}}\left(P\leq\tau\right)>0$, (ii) $UNI[0,1]\leq_{st}P^{(\theta^{\star})}$, and (iii)  $P^{(\theta^{\star})}\leq_{\text{rh}}P^{(\theta)}$ for all $\theta \in \mathcal{M}$, where $P^{(\tilde{\theta})}$ is a random variable possessing the distribution of $P$ under $\tilde{\theta}$.  

Then, 
\begin{equation*}
\frac{P}{\mathbb{P}_{\theta^{\star}}\left(P\leq\tau\right)}\;\;\text{given}\;\;P\leq\tau
\end{equation*}
is stochastically not smaller than $UNI[0,1]$, under any $\theta\in \mathcal{M}$.
\end{lemma}

\begin{proof}
For $\theta\in\mathcal{M}$, we have to show that
\begin{equation}\label{eq1}
    \mathbb{P}_{\theta}\big(P\leq t\mathbb{P}_{\theta^{\star}}\left(P\leq\tau\right)\big)\leq t\mathbb{P}_{\theta}\big(P\leq \mathbb{P}_{\theta^{\star}}\left(P\leq\tau\right)\big)
\end{equation}
holds true for all $t\in[0,1]$. To this end, notice that
we assumed that $\mathbb{P}_{\theta}(P\leq t)/\mathbb{P}_{\theta^{\star}}(P\leq t)$ is non-decreasing in $t$. Thus, it holds 

\begin{equation*}
    \frac{\mathbb{P}_{\theta}\big(P\leq t\mathbb{P}_{\theta^{\star}}\left(P\leq\tau\right)\big)}{\mathbb{P}_{\theta^{\star}}\big(P\leq t\mathbb{P}_{\theta^{\star}}\left(P\leq\tau\right)\big)}\leq \frac{\mathbb{P}_{\theta}\big(P\leq \mathbb{P}_{\theta^{\star}}\left(P\leq\tau\right)\big)}{\mathbb{P}_{\theta^{\star}}\big(P\leq \mathbb{P}_{\theta^{\star}}\left(P\leq\tau\right)\big)},\;\;t\in[0,1].
\end{equation*}

With $\mathbb{P}_{\theta^{\star}}\big(P\leq \mathbb{P}_{\theta^{\star}}\left(P\leq\tau\right)\big)=\mathbb{P}_{\theta^{\star}}\left(P\leq\tau\right)$ and $\mathbb{P}_{\theta^{\star}}\big(P\leq t\mathbb{P}_{\theta^{\star}}\left(P\leq\tau\right)\big)\leq t\mathbb{P}_{\theta^{\star}}\left(P\leq\tau\right)$, the proof is complete.
\end{proof}

We now establish the validity of the randomized $p$-values defined above.

\begin{theorem}[Validity of the proposed  $p$-values]\label{s:validity}
Under our general assumptions specified in Section \ref{general-assum}, assume that $(P_\theta)_{\theta \in \Theta}$ constitutes a one-parametric exponential family in natural parametrization, and that the test statistic $T(\pmb{X})$ is the natural sufficient statistic for $\theta$ in the product model given by $(\mathbb{P}_\theta)_{\theta \in \Theta}$. 

Then, the LFC-based $p$-value $P^{LFC}(\pmb{X})$, the single-stage randomized $p$-value $P^{rand1}(\pmb{X},U,c)$, the UMP $p$-value $P^{rand}_T(\pmb{X},U)$, and the two-stage randomized $p$-value
$P^{rand2}(\pmb{X},U,\tilde{U},c)$ are all valid $p$-values, for all $c\in[0,1]$. 
\end{theorem}

\begin{proof} Throughout the proof, we use the closure property for an exponential family of distributions (cf. Theorem $18.5$ in \cite{dasgupta2011probability}, page $595$). This property yields that  the distribution of $T(\pmb{X})$ belongs to the same type of exponential family of distributions as that of $\pmb{X}$. Such a distribution is known to possess a monotone likelihood ratio with respect to $\theta$ (cf. \cite{karlin1956distributions}).

\begin{enumerate}
\item[1.)]
We first show that $P^{LFC}(\pmb{X})$ is a uniformly valid (hence valid) $p$-value, by showing that it fulfils the three conditions in Lemma~\ref{s:validityrand1}, for $\mathcal{M} = H$. To this end, let $\tau \in (0, 1]$ fulfilling $\mathbb{P}_{\theta^{\star}}\left(P^{LFC}(\pmb{X})\leq\tau\right)>0$ be arbitrary, but fixed. Such a $\tau$ is guaranteed to exist, because $\mathbb{P}_{\theta^{\star}}\left(P^{LFC}(\pmb{X})\leq 1 \right) = 1 >0$, and it holds that $
    \mathbb{P}_{\theta^{\star}}\big(P^{LFC}(\pmb{X})\leq \mathbb{P}_{\theta^{\star}}\left(P^{LFC}(\pmb{X})\leq\tau\right)\big)=\mathbb{P}_{\theta^{\star}}\left(P^{LFC}(\pmb{X})\leq\tau\right)$. Hence, condition (i) in Lemma \ref{s:validityrand1} is fulfilled.  The validity of condition (ii) immediately follows from the principle of quantile and probability integral transforms; see, e.\ g., Section 2.2 in \cite{dickhaus-nonparametric-tests}. Finally, we establish the validity of condition (iii) in Lemma \ref{s:validityrand1} by noticing that $(P^{LFC}(\pmb{X}))^{(\theta^{\star})}\leq_{\text{rh}}(P^{LFC}(\pmb{X}))^{(\theta)}$ holds true for all $\theta\in H$. The last statement follows from the fact that, within an exponential family, the distributions of the test statistic $T(\pmb{X})$ are monotone likelihood ratio ordered with respect to the parameter $\theta$, and therefore so are the distributions of $P^{LFC}(\pmb{X})=1-F_{\theta^{\star}}(T(\pmb{X})-)$. Utilizing the definition of the reverse hazard rate order, we obtain that
\begin{equation*}
\hspace{70pt minus 1fil}
\frac{P^{LFC}(\pmb{X})}{\mathbb{P}_{\theta^{\star}}\left(P^{LFC}(\pmb{X})\leq\tau\right)}\;\;\text{given}\;\;P^{LFC}(\pmb{X})\leq\tau
\end{equation*}
is stochastically not smaller than $UNI[0,1]$, for any $\theta\in H$. Since $\tau$ has been chosen arbitrarily,  we conclude that $P^{LFC}(\pmb{X})$ is uniformly valid.

\item[2.)]
To show that $P^{rand1}(\pmb{X},U,c)$ is valid, we exploit that its non-randomized version $P^{LFC}(\pmb{X})$ is uniformly valid as shown in (1.),  and we calculate that
\begin{eqnarray*}
\mathbb{P}_{\theta}\left(P^{rand1}(\pmb{X},U,c)\leq t\right)&=&\mathbb{P}_{\theta}\left(U\leq t\right)\mathbb{P}_{\theta}\left(P^{LFC}(\pmb{X})\geq c\right) + \mathbb{P}_{\theta}\left(P^{LFC}(\pmb{X})\leq tc^{\star}\right)\\
&\leq&t\mathbb{P}_{\theta}\left(P^{LFC}(\pmb{X})\geq c\right) + t\mathbb{P}_{\theta}\left(P^{LFC}(\pmb{X})\leq c^{\star}\right) = t,
\end{eqnarray*}
where in the last step we used Lemma~\ref{s:validityrand1}, Eq. \eqref{eq1}, and the definition of $c^{\star} =\mathbb{P}_{\theta^*}\left(P^{LFC}(\pmb{X})<c\right)$.


\item[3.)]
In order to show that $P^{rand2}(\pmb{X},U,\tilde{U},c)$ which is based on $P_T^{rand}(\pmb{X},U)$ is a valid $p$-value, we show that 
 condition $(1.)$ in Theorem$~1$ of \cite{hoang2021usage} is fulfilled for $P_T^{rand}(\pmb{X},U)$. It is therefore sufficient to show that 
\begin{equation}\label{tc-ineq}
\mathbb{P}_\theta\{P_T^{rand}(\pmb{X},U)\leq tc\}\leq t\mathbb{P}_\theta\{P_T^{rand}(\pmb{X},U)\leq c\}
\end{equation}
holds for all $c\in [0,1]$, $t\in [0,1]$, and all $\theta\in H$. In the case of $tc = 0$, \eqref{tc-ineq} is trivially fulfilled. In the case of $tc > 0$, \eqref{tc-ineq} is equivalent to 
\[\hspace{70pt minus 1fil} \frac{\mathbb{P}_\theta\{P_T^{rand}(\pmb{X},U)\leq tc\}}{tc}\leq\frac{\mathbb{P}_\theta\{P_T^{rand}(\pmb{X},U)\leq c\}}{c}\]
and therefore to 
\[\hspace{70pt minus 1fil} \frac{\mathbb{P}_\theta\{P_T^{rand}(\pmb{X},U)\leq tc\}}{\mathbb{P}_{\theta^*}\{P_T^{rand}(\pmb{X},U)\leq tc\}}\leq\frac{\mathbb{P}_\theta\{P_T^{rand}(\pmb{X},U)\leq c\}}{\mathbb{P}_{\theta^*}\{P_T^{rand}(\pmb{X},U)\leq c\}}  \]
holding for all $c\in [0,1]$, $t\in [0,1]$, and all $\theta\in H$. Compare this to the definition of the reverse hazard rate order. For a parameter value $\theta$ in the null hypothesis, i.e. $\theta\leq \theta^*$, we define the function 
\[\hspace{90pt minus 1fil} h(t)=\frac{\mathbb{P}_{\theta}\{P_T^{rand}(\pmb{X},U)\leq t\}}{\mathbb{P}_{\theta^*}\{P_T^{rand}(\pmb{X},U)\leq t\}}.
\]
It remains to show that $h$ is monotonically increasing in $t$. Let $S$ be the set of support points of $P^{LFC}(\pmb{X})$. The CDFs of $P_T^{rand}(\pmb{X},U)$ and $P^{LFC}(\pmb{X})$ coincide on elements in $S$. For two points $t_1, t_2\in S$, if $t_1<t_2$ then $h(t_1)\leq h(t_2)$, since, as elaborated above, the distributions of $P^{LFC}(\pmb{X})$ are reverse hazard rate ordered. Both the numerator and denominator of $h(t)$ are linear functions in $t$ between $t_{1}$ and $t_{2}$. 
Let $a_1$, $a_2$, $b_1$ and $b_2$ be positive constants such that $h(t)=\frac{a_1t+b_1}{a_2t+b_2}$, $t\in[0,1]$. For example, define
\[\hspace{70pt minus 1fil} b_{1}=\mathbb{P}_{\theta}\{P_T^{rand}(X,U)\leq t_{1}\}=\mathbb{P}_{\theta}\{P^{LFC}(X)\leq t_{1}\}\] and 
\[\hspace{60pt minus 1fil} b_{1}+a_{1}=\mathbb{P}_{\theta}\{P_T^{rand}(X,U)\leq t_{2}\}=\mathbb{P}_{\theta}\{P^{LFC}(X)\leq t_{2}\}.\]
It then holds $\frac{b_1}{b_2}\leq \frac{a_1+b_1}{a_2+b_2}$, $a_1a_2^{-1},b_1b_2^{-1}\geq 1$, and that the function $t\mapsto \frac{a_1t+b_1}{a_2t+b_2}$, for $t\in [0,1]$, is increasing in $t$, completing the argumentation.

\item[4.)]
To show that $P_T^{rand}(\pmb{X},U)$ is a valid $p$-value, we need to show that $\mathbb{P}_{\theta}(P_T^{rand}(\pmb{X},U)\leq t)\leq t$ 
holds true for all $t\in [0,1]$.
To this end, notice that
\begin{equation*}
\hspace{70pt minus 1fil} \mathbb{P}_\theta\{P_T^{rand}(\pmb{X},U)\leq t\tau\}\leq t\mathbb{P}_\theta\{P_T^{rand}(\pmb{X},U)\leq \tau\}
\end{equation*}
holds true for all $\tau\in (0,1]$, $t\in [0,1]$, and all $\theta\in H$ as shown in (3.) above. Setting $\tau=1$, this results in 
$\mathbb{P}_\theta\{P_T^{rand}(\pmb{X},U)\leq t\}\leq t$, which is what we needed to show.
\end{enumerate}
\end{proof}

With the randomized $p$-values so defined and their validity proved, we are now in a position to apply them to the general binomial model and later in the estimation of proportion of true null hypotheses.


\section{APPLICATION TO THE BINOMIAL MODEL}
\label{s:binomial}
\subsection{General properties}
In this section, we demonstrate how the proposed randomized $p$-values can be used when the data are constituted by Bernoulli indicators. We denote the binomial distribution with parameters $n \in \mathbb{N}$ (number of trials) and $\theta \in [0, 1]$ (success probability) by $Bin(n,\theta)$. In the case of $n=1$, we have the Bernoulli distribution with success parameter $\theta$, Bernoulli$(\theta)$ for short. To avoid pathologies, we will restrict attention to $\theta \in (0, 1)$ at some occasions. Suppose we are interested in testing the hypothesis $H: \theta\leq \theta^*$ versus $K: \theta> \theta^*$, where  $\theta^*$ is a pre-specified constant. Let $\pmb{X} =(X_1,\ldots,X_n)^\top$ be i.i.d., with $X_1 \sim \text{Bernoulli}(\theta)$. The test statistic $T(\pmb{X})=\sum_{i=1}^{n} X_i$ follows the $Bin(n,\theta)$ distribution, and it tends to larger values under the alternative. Throughout this section, we let $\mathbb{P}_{\theta}$ denote the $n$-fold product of Bernoulli$(\theta)$ and we let $F_{\theta}$ denote the CDF of $Bin(n,\theta)$. 

\begin{figure}[htb]
\begin{center}
\includegraphics[width=6 in]{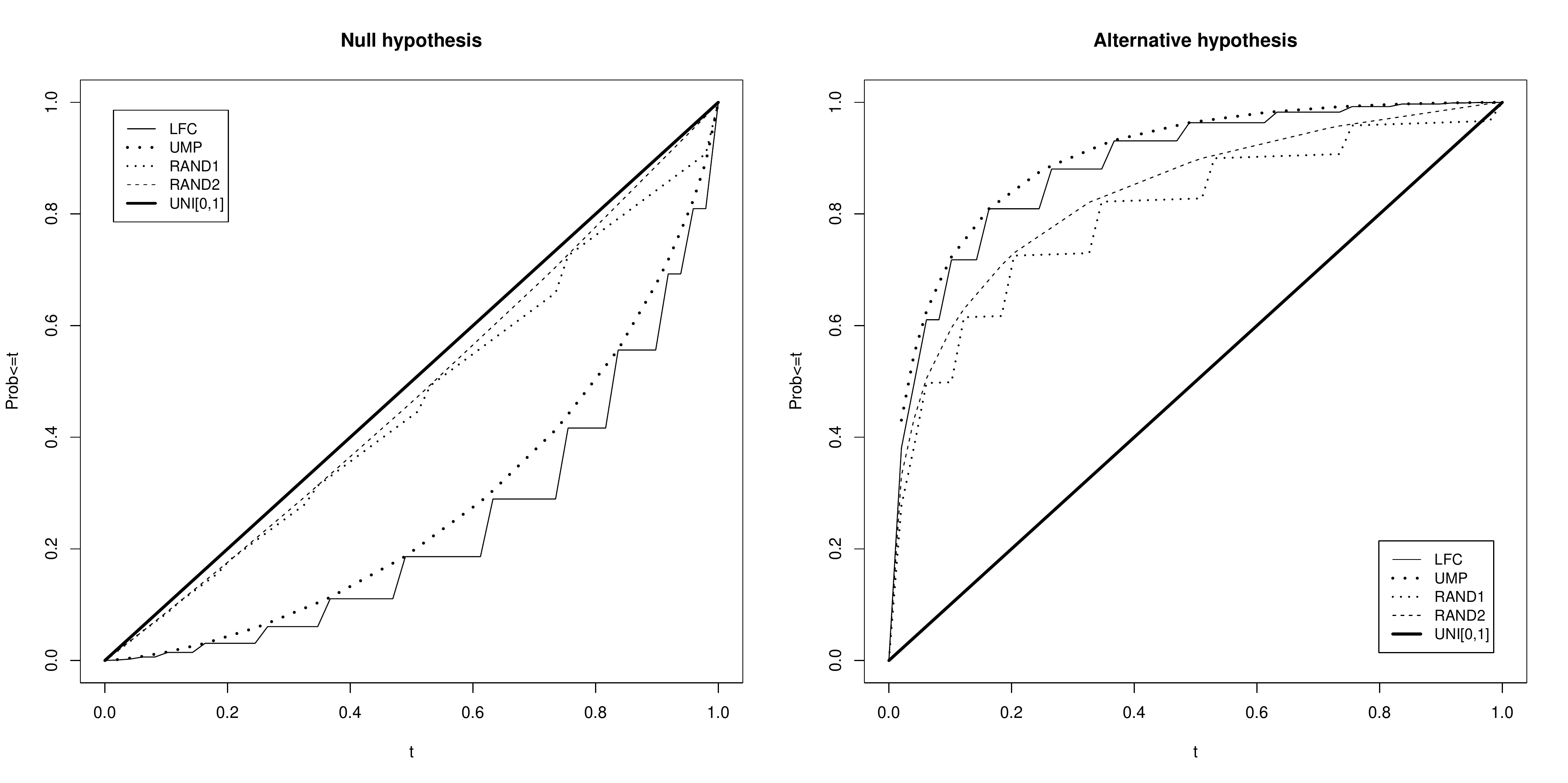}
\caption{ An illustration of the CDF under null and alternative hypothesis for LFC, UMP, single-stage randomized $p$-value (RAND1), and two-stage randomized $p$-value (RAND2) for $n=50$, $\theta^*=0.25$, and $c=0.5$.}
\label{f:fig1}
\end{center}
\end{figure}

With these specifications, the LFC-based $p$-value under the binomial model is given by 
$\eqref{eq:lfc1}$, and the corresponding CDF for $p$-LFC is given by $\eqref{eq:lfc3}$. The single-stage randomized $p$-value is given by $(\ref{eq:r4})$ using the LFC-based $p$-value described in the first sentence of this paragraph. Equation $(\ref{eq:r5})$ gives the CDF for the single-stage randomized $p$-value under the binomial model.
Turning our attention to the two-stage randomization procedure, the randomized $p$-value in the first stage for the binomial model is given by $(\ref{eq:r1})$ with the corresponding CDF given by $\eqref{eq:w1}$. Similarly, the randomized $p$-value in the second stage for the binomial model is given by $(\ref{eq:r2})$ with the corresponding CDF given by
$\eqref{eq:r3}$. Figure $\ref{f:fig1}$ displays the CDFs of the LFC-based $p$-value (denoted by LFC), the UMP $p$-value $P_T^{rand}$ (denoted by UMP), the single-stage randomized $p$-value (denoted by RAND1), and the two-stage randomized $p$-value (denoted by RAND2) under the null ($\theta=0.20$) and under the alternative hypothesis ($\theta=0.37$), respectively, where $n=50$, $\theta^*=0.25$, and  $c=0.5$.

Figure $\ref{f:fig1}$ indicates that under the alternative hypothesis the test based on $P^{rand1}$ is the least powerful one, followed by that based on $P^{rand2}$. On the other hand, $P^{rand2}$ is the least conservative of the considered $p$-values under the null hypothesis, meaning that its CDF is closest to the main diagonal in the unit square. The latter property is very useful in situations where closeness of the $p$-value distribution to UNI$[0, 1]$ under the null is required. For example, this is the case when estimating of proportion of true null hypotheses by means of an estimator which is based on the empirical CDF of all $p$-values in a multiple testing context; see \cite{dickhaus2013randomized} and \cite{hoang2021usage}. We will return to this estimation problem in Section \ref{s:multiple}. 

The $p$-value $P^{rand1}$ is partially discrete as can be seen by the slanting steps in its CDF under both the null and alternative hypothesis. We also note that $P^{rand2}(\pmb{X},U,\tilde{U},1)=P_{T}^{rand}(\pmb{X},U)$, which is always stochastically smaller than $p$-LFC.  The distribution of $P^{rand2}$ moves closer to UNI$[0,1]$ distribution if $c\to 0$. However, for $c$ less but close to $1$, the test based on $P^{rand2}$ can still be more powerful than that based on $p$-LFC for many significance levels $\alpha\in(0,1)$.


\subsection{Sample size versus power}

We now investigate the relationship between the powers of the tests based on the four considered $p$-values (UMP, LFC, $P^{rand1}$, and $P^{rand2}$) and an increasing sample size. In comparing the power of the tests based on the $p$-values with different sample sizes, we are particularly interested in assessing whether there is an increase in power with an increase in sample size, as one would (usually) expect. When testing for a binomial parameter in the composite null hypothesis, \cite{finner2001ump} showed that the power function at a sample size $n$ can be higher than at some $n+i$, $i\geq 1$ for the non-randomized $p$-value (LFC-based $p$-value), which seems paradoxical. 
This paradox also occurs in permutation tests, in Fisher's exact test, and when comparing ratios or differences between two binomial success parameters. This drop in power when the sample size increases slightly can occur since power depends on the actual alpha rather than the nominal alpha. An increased sample size leads to a decreased actual alpha. The drop in power is also caused by the discrete nature of the test statistic involved in the definition of the LFC-based $p$-value. 

\begin{figure}
\centerline{\includegraphics[width=5in]{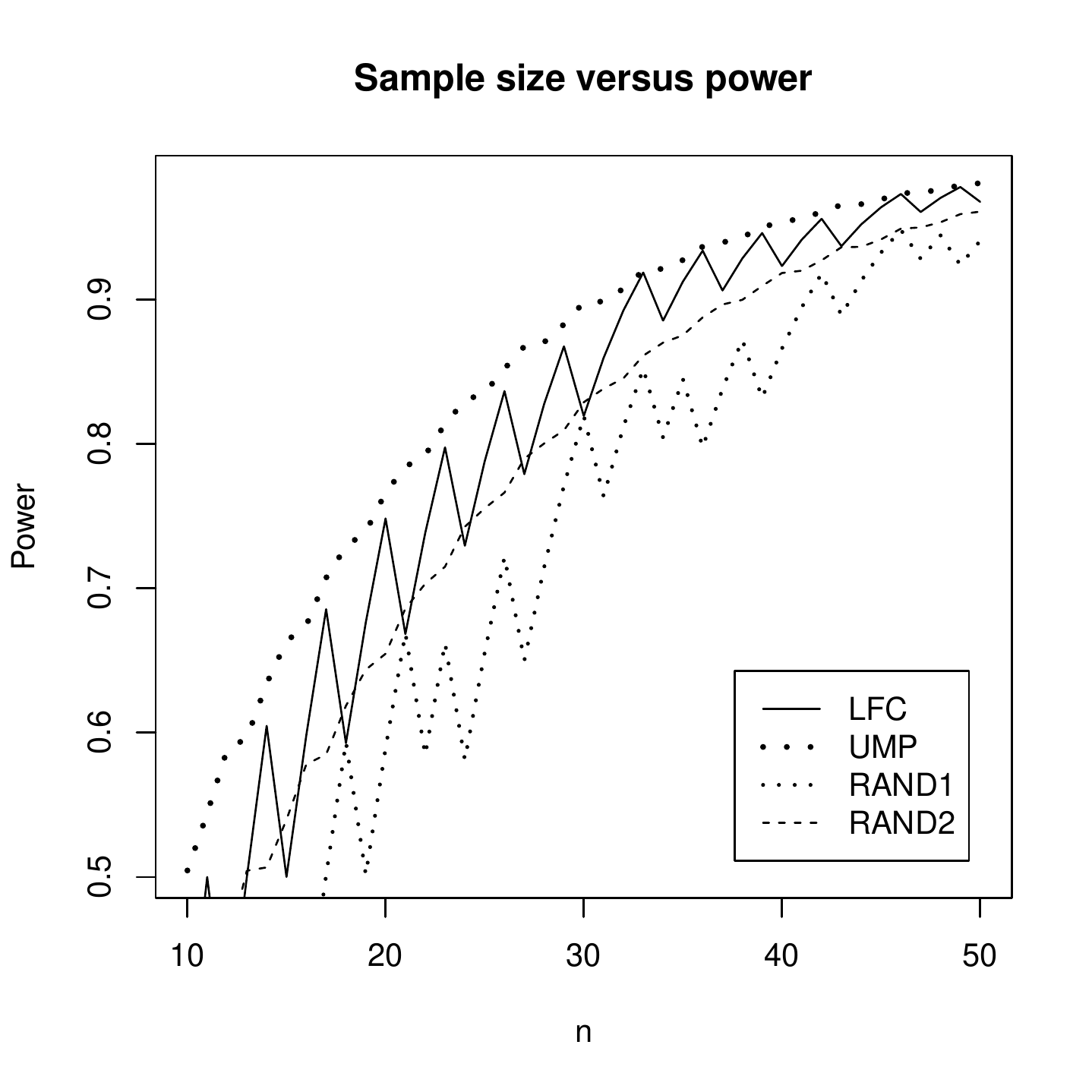}}
\caption{Graphic display of power versus sample size for LFC, UMP, single-, and two-stage randomized $p$-values (RAND1) and (RAND2), respectively, for significance level $\alpha=0.05$, $\theta^*=0.25$, and $c=0.5$.}
\label{f:fig2}
\end{figure}

In Figure $\ref{f:fig2}$, we compare the powers of the tests based on the LFC-based $p$-value, the UMP $p$-value and the randomized $p$-values $P^{rand1}$ and $P^{rand2}$ for different sample sizes, $\theta^*=0.25$, $\theta=0.5$, $c=0.5$, and $\alpha=0.05$.
Indeed, Figure $\ref{f:fig2}$ demonstrates that the power of the test based on $p$-LFC is not monotonically increasing in the sample size. The power function of the test based on $P^{rand1}$ shows a similar behaviour to the one based on $p$-LFC. Instead, the power function is monotonically increasing in $n$ when utilizing $P^{rand2}$ or $P_T^{rand}$. Of course, the test based on the UMP $p$-value $P_T^{rand}$ is the most powerful of all the tests based on the four $p$-values.
The $p$-value $P^{rand1}$ is based on $p$-LFC, while $P^{rand2}$ is based on the stochastically smaller $P_T^{rand}$. Therefore, it is expected  that the test based on $P^{rand2}$ is also more powerful and less conservative than that based on $P^{rand1}$. This behavior is verified in Figures  $\ref{f:fig1}$ and $\ref{f:fig2}$.

\subsection{Power and level of conservativeness for different values of \texorpdfstring{$c$}{c}}

\begin{figure}
\centerline{\includegraphics[width=6.5in]{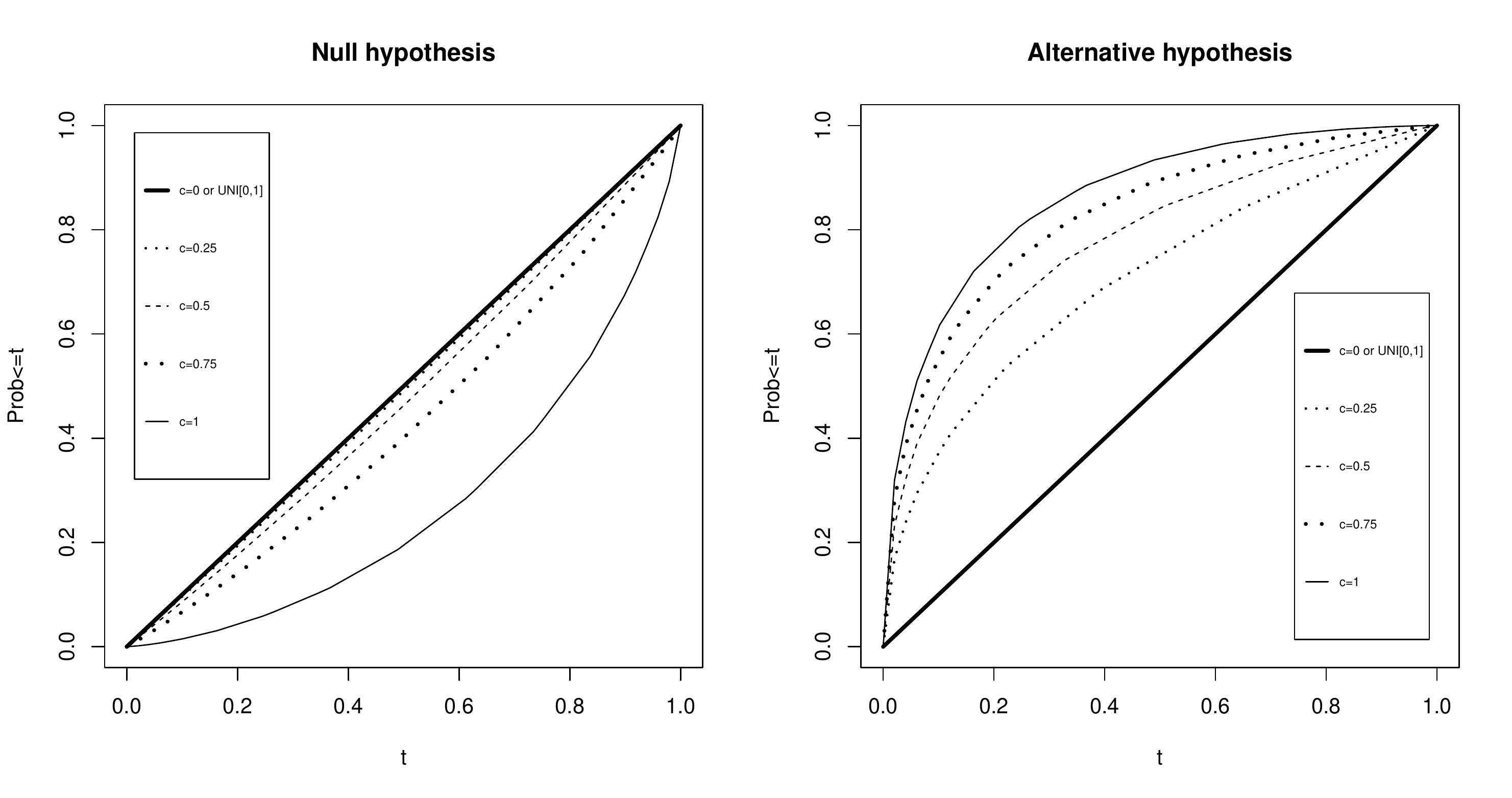}}
\caption{An illustration of the CDF for two-stage randomized $p$-value (RAND2) in the null ($\theta=0.20$) and alternative ($\theta=0.35$) hypotheses under different values of $c$ for $n=50$ and $\theta^*=0.25$.}
\label{f:fig3}
\end{figure}

\begin{figure}
\centerline{\includegraphics[width=4.6in]{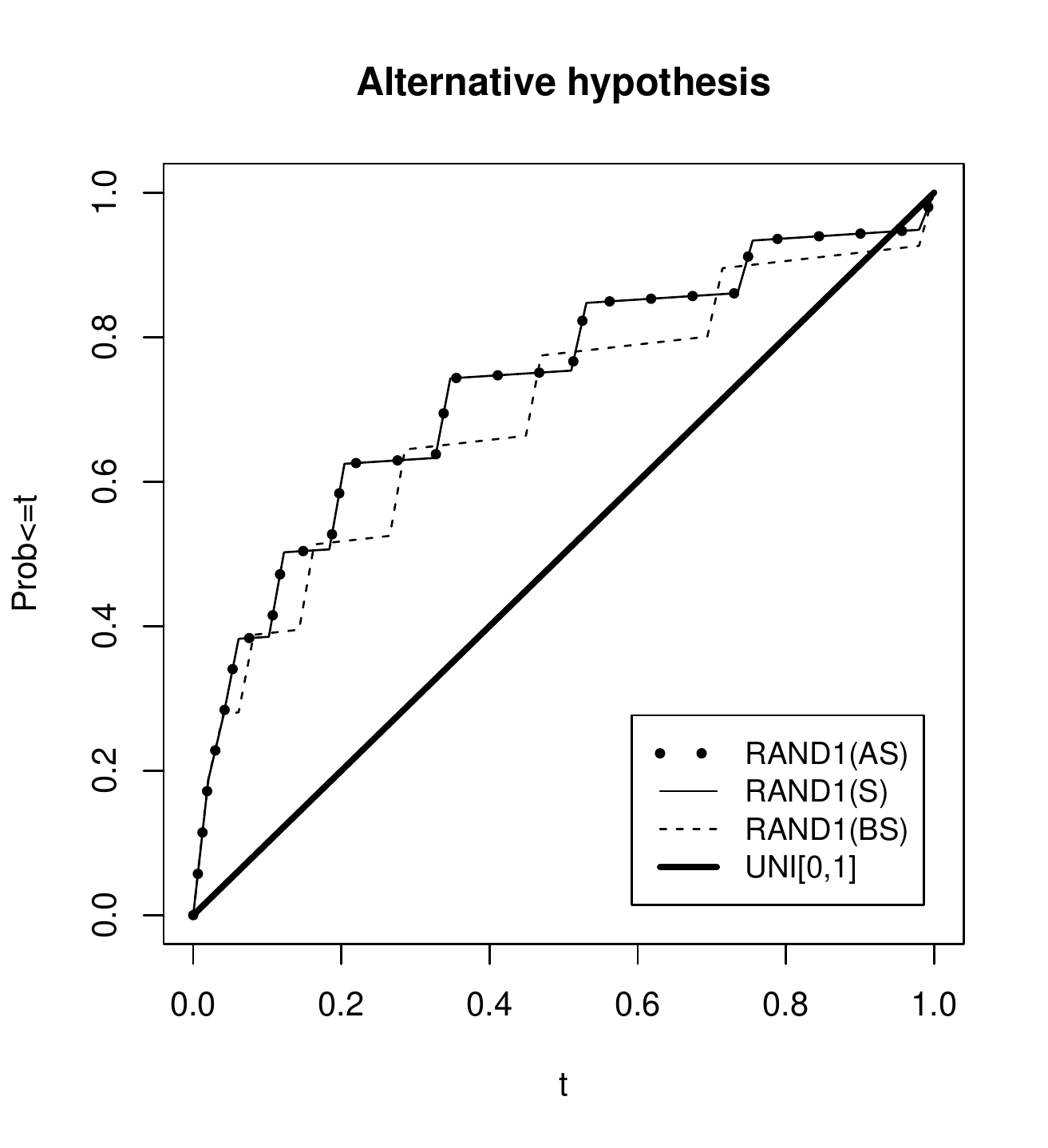}}
\caption{Graphic illustration of the CDF for single-stage randomized $p$-value in  support-RAND1(S), below support-RAND1(BS), and above support-RAND1(AS) for $n=50$, $\theta^*=0.25$, $\theta=0.35$, and $c=0.5$.}
\label{f:fig4}
\end{figure}

We now investigate the CDFs for the two randomized $p$-values $P^{rand1}$ and $P^{rand2}$ when different values of $c$ are used. In Figure $\ref{f:fig3}$, we display  the CDFs for $P^{rand2}$ under the null and alternative hypothesis. The power function of the test based on $P^{rand1}$ when $c^*$ (which is a support point for $p$-LFC) is used is displayed in  Figure $\ref{f:fig4}$. Under the null hypothesis in Figure $\ref{f:fig3}$, as $c$ increases, the CDF departs from the diagonal line ($c=0$) which is the least conservative, to the CDF of the UMP $p$-value ($c=1$) which is the most conservative. Under the alternative hypothesis, the pointwise largest CDF occurs when $c=1$, and the pointwise smallest CDF occurs when $c=0$. 

In Figure $\ref{f:fig4}$, the power function of the test based on $P^{rand1}(\pmb{X},U,c^*)$  is displayed. We indicate the graph of this function by RAND1(S) in Figure $\ref{f:fig4}$. In addition, we display in Figure $\ref{f:fig4}$ two further curves, indicated by RAND1(AS) and RAND1(BS). These are the graphs of the power functions of the test based on the single-stage randomized $p$-values when $c^*+\varepsilon$ or $c^*-\varepsilon$ are used, respectively, where $\varepsilon$ is small.

For any support point $c^*$, RAND1(AS) and RAND1(S) are equal as explained after (\ref{eq:r5}). The $p$-values RAND1(S) and RAND1(BS) have different CDFs regardless of the size of $\varepsilon$ since the support point in RAND1(BS) is always smaller than the support point $c^*$ in RAND1(S). Unlike $P^{rand2}$, the single-stage randomized $p$-value $P^{rand1}$  is not stochastically ordered in $c$.

\begin{example}[Application in Group Testing]\label{grouptest}
In this example, we  illustrate how the four $p$-values $p$-LFC, $P^{rand1}$, UMP, and $P^{rand2}$ can be used in group testing. Assume that we have a population of size $N$ that can be divided into $g$ groups, each of size $s$. The individuals in each group are either positive (denoted by 1) or negative (denoted by 0) with respect to the target event, and this outcome is random. To take care of the possibility of misclassification for the groups due to errors in testing, define an indicator function $Q_{\ell}=1$ if the $\ell$-th group is truly positive with $P(Q_{\ell}=1)=1-(1-\theta)^s$ and $Q_{\ell}=0 $ otherwise with $P(Q_{\ell}=0)=(1-\theta)^s$, and another indicator function $T_{\ell}=1$ if the $\ell$-th group tests positive and $T_{\ell}=0$ otherwise, for $\ell=1,\ldots,g$. In this, we call a group positive  if it contains at least one positive individual, and negative otherwise. Define the group sensitivity $S_e=P(T_{\ell}=1|Q_{\ell}=1)$ and specificity  $S_p=P(T_{\ell}=0|Q_{\ell}=0)$. We  assume that the trait under consideration and the sampling scheme are such, that $Q_1,\ldots,Q_g$ are i.i.d. Bernoulli random variables, each with mean $\pi$, where $\pi=S_e+(1-S_e-S_p)(1-\theta)^{s}$. Letting $R=\sum_{\ell=1}^{g} Q_{\ell}$ denote the total number of positive groups, we have that $R\sim Bin(g,\pi)$. A realization of $R$ will be denoted by $r$. The maximum likelihood estimator (MLE) for $\theta$ is $\widehat{\theta}=1-\{(\widehat{\pi}-S_e)/(1-S_e-S_p)\}^{1/s}$, where $\widehat{\pi}=r/g$ is the proportion of positive groups out of the $g$ groups. The pair of composite hypotheses $H: \theta\leq \theta^*$ versus $K: \theta>\theta^*$ considered in Section $\ref{s:binomial}$ is similar to the case for individual inspection without misclassification, that is, when $S_e=S_p=s=1$. This pair of hypotheses is equivalent to $H_s: \pi \leq \pi^*$ versus $K_s: \pi>\pi^*$ for group testing where $\pi^*= S_e+(1-S_e-S_p)(1-\theta^*)^{s}$ and $\pi=S_e+(1-S_e-S_p)(1-\theta)^{s}$. We consider a test statistic  $T = T(\pmb{Q})$, where $\pmb{Q}=(Q_{1},\ldots,Q_{g})^{\top}$. The computations for the CDFs of the four $p$-values for the hypothesis $H_s: \pi \leq \pi^*$ versus $K_s: \pi>\pi^*$ follow the same process as in Section $\ref{s:binomial}$, but with $\theta^*$,\ $\theta$, and $n$ replaced by $\pi^*$, \ $\pi$,\ and $g$, respectively and the sensitivity and specificity can be taken to be a constant, for example, $S_e=S_p=0.95$. We compare the power of the tests based on the four $p$-values for different group sizes $s$ under two cases, namely (i) for a fixed number of inspections, and (ii) for a fixed number of individuals. A brief summary of the two cases is given in (iii).


\begin{enumerate}
\item [(i)]Fixed number of inspections:
 This is common, for example, in multiple-vector transfer designs where the number of inspections is limited by the available number of plants, green house spaces, or isolation cages for the plants. However, any number of insects can be used since the cost of obtaining an insect is small, see \cite{swallow1985group}, \cite{tebbs2003more}, and \cite{mccann2007pairwise} for more details. It is expected that inspecting items in groups when the number of inspections is fixed will improve the power of the tests based on the four $p$-values compared to performing (the same number of) individual inspections. It is also expected that the power of the tests based on the four $p$-values will further increase with an increase in group size as long as the proportion of positive individuals in a group is low. We give an illustration for this case in Figure $\ref{fig5}$.

\begin{figure}
\begin{center}
\includegraphics[width=4.6in]{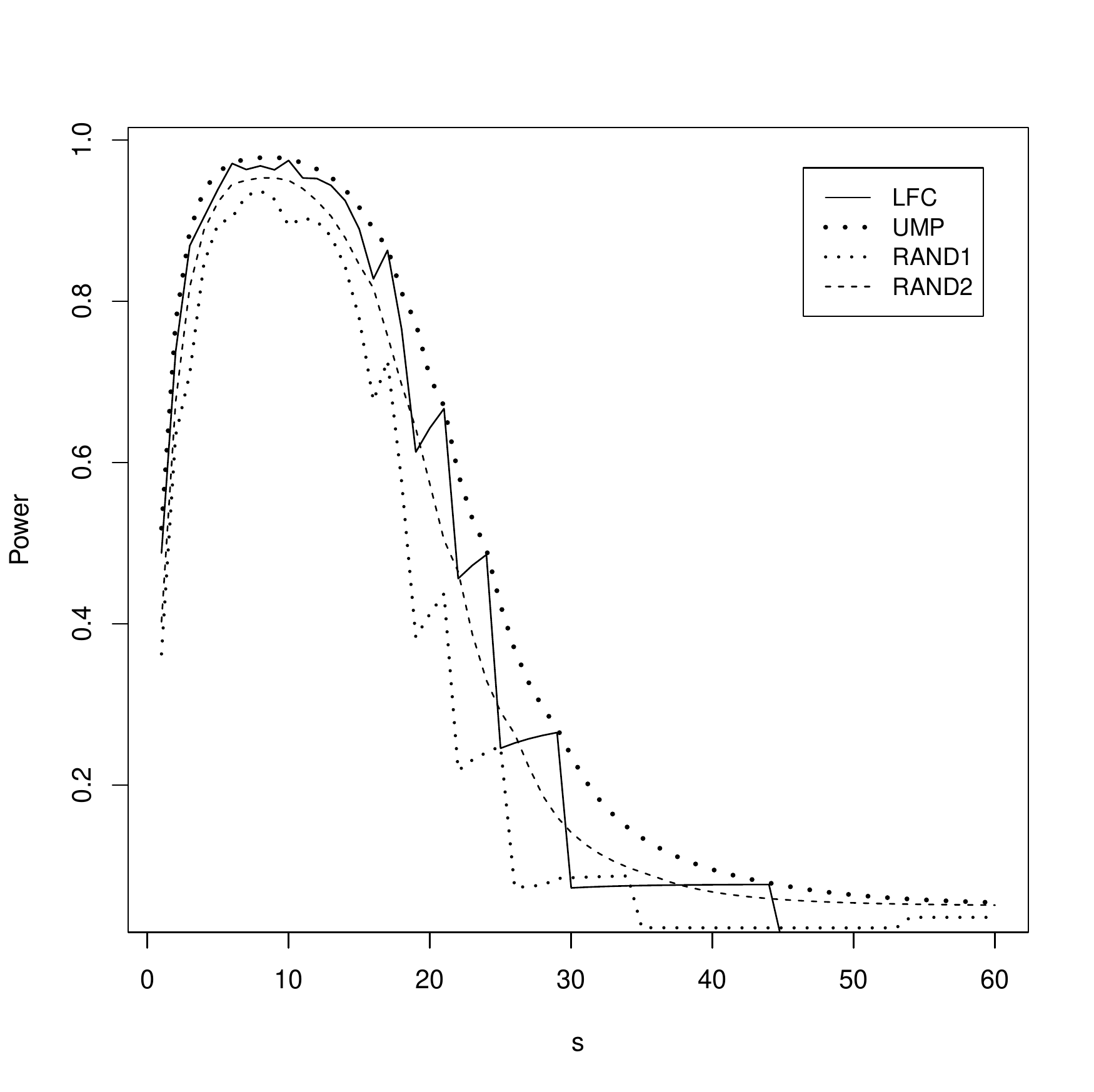}
\caption{An illustration of power versus different group sizes $s$ for LFC, UMP single- (RAND1) and two-stage (RAND2) randomized $p$-values for fixed number of tests. We set $c=0.5$, $\alpha=0.05$, $g=50$, $\theta^*=0.1$, $\theta=0.2$, and  $S_e=S_p=0.95$.}
\label{fig5}
\end{center}
\end{figure}

Increasing the group size $s$ increases the power first, then eventually the power drops towards zero for large group sizes $s$ and a large prevalence rate as illustrated in Figure $\ref{fig5}$. The aforementioned drop in power occurs since $\pi^*$ increases in $s$, and for $s$ too large $K_s:\pi>\pi^*$ becomes too difficult to detect. 
Generally for a high prevalence rate $\theta$, the power functions increase faster and drop earlier compared to a low prevalence rate. For a large $\theta^*$ the power functions drop earlier compared to a low $\theta^*$. Of course, the power of the test based on the UMP $p$-value is the highest (among the four tests). The test based on $P^{rand1}$ has the lowest power throughout while the tests based on $p$-LFC and $P^{rand2}$ are competing. The optimal group size based on Figure $\ref{fig5}$ is in the vicinity of $s=10$.

\item [(ii)] Fixed number of individuals:
We can also have a situation where the total number of individuals is fixed but the number of groups is not fixed. We give a display in Figure $\ref{fig6}$ of how the power of the tests based on all the four $p$-values decreases with an increase in group size. We set $\theta^*=0.1$, $\theta=0.2$, $c=0.5$, $\alpha=0.05$, $S_e=S_p=0.95$, and $s*g=300$. The power of the tests based on $P^{rand2}$ and UMP $p$-value drops to $\alpha$ and stays that way throughout never dropping to zero. The CDF of $P^{rand2}$ lies between the CDF of $P_T^{rand}$ and the CDF of UNI$[0,1]$, cf. \cite{hoang2021usage}, which implies that when the CDF of $P_T^{rand}$ at $\alpha$ goes to $\alpha$ then the CDF of  $P_T^{rand}$ at $\alpha$ will also converge to $\alpha$. 

\begin{figure}[htb]
\begin{center}
\includegraphics[width=4.6in]{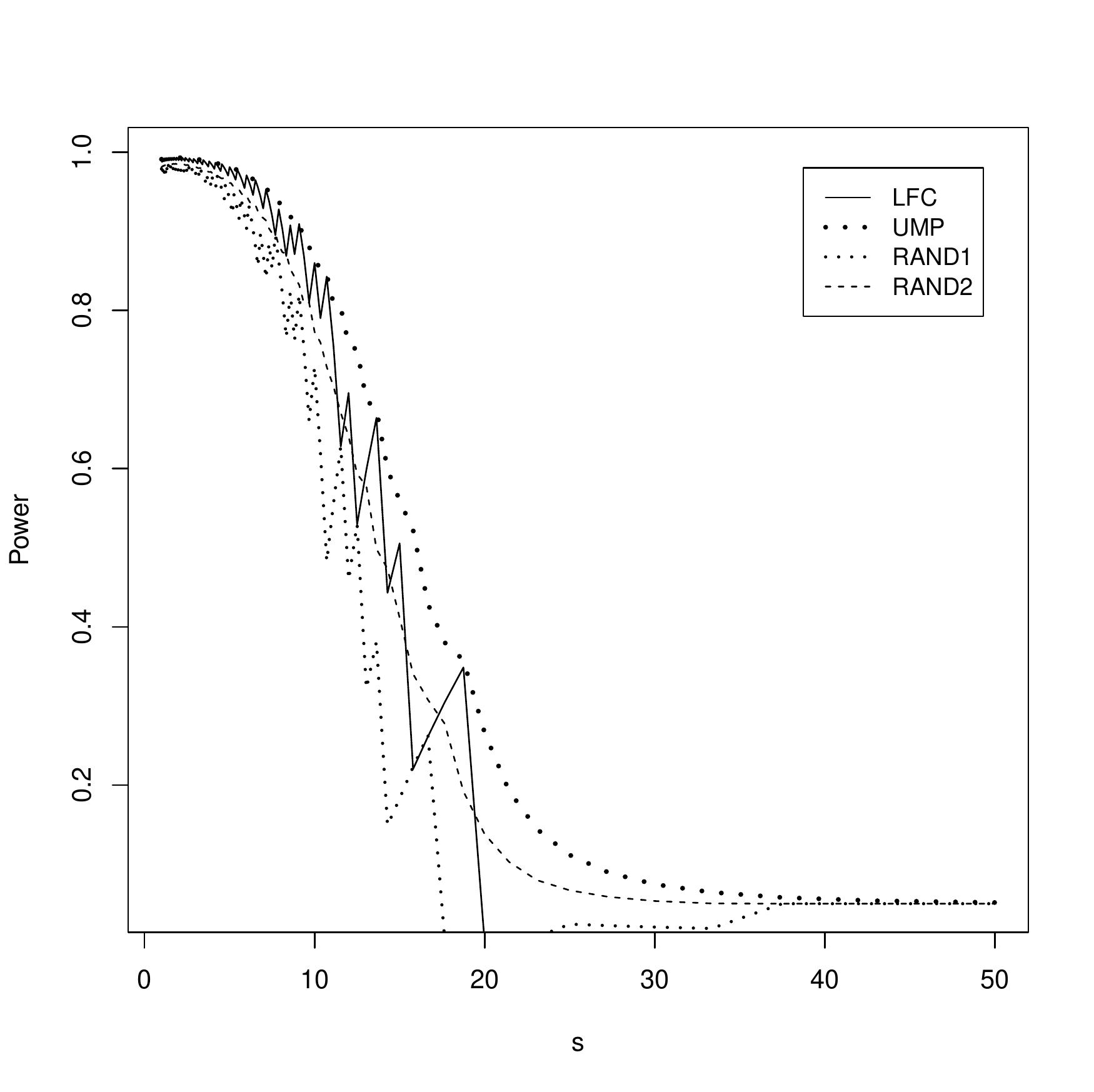}
\caption{An illustration of power versus different group sizes $s$ for LFC, UMP single- (RAND1) and two-stage (RAND2) randomized $p$-values for fixed number of individuals. We set $c=0.5$, $\alpha=0.05$, $\theta^*=0.1$, $\theta=0.2$, $N=300$, $S_e=S_p=0.95$, $g$ is taken to be a decreasing sequence of integers from $300$ to $6$, and $s=N/g$. The non-integers in $s$ are used the way they are.}
\label{fig6}
\end{center}
\end{figure}

The tests based on all the four $p$-values have a higher power when doing individual inspection $(s=1)$ compared to the case of group inspection $(s>1)$. This is because individual inspection provides the maximum amount of information when the number of individuals is fixed and hence a higher power than when inspecting individuals in groups (\cite{tebbs2003more}). The power gained by the tests in the case of individual inspection is not much and may not justify the additional inspection costs. For example, when $\theta=0.008$, under individual inspection, the power of tests based on $p$-LFC is $69.3\%$ and the maximum number of inspections, in this case $300$,  is required. At the same prevalence when using group testing with $s=10$, the power of tests based on $p$-LFC is $68.5\%$  and only $30$ inspections are used. In this example, the slight gain in power by the tests based on $p$-LFC under individual inspection requires ten times more inspections compared to when the same $p$-value is used in group testing. Therefore, in applications where the inspection budget is limited, group testing  will still remain very useful despite the slight loss in power.

\item [(iii)] A summary of the two cases:
In both cases considered above, for large group sizes ($s\geq 20$) as seen in Figures $\ref{fig5}$ and $\ref{fig6}$, the power of the tests based on $p$-LFC and $P^{rand1}$ drops to zero, the one for $P^{rand1}$ rises again to $0.05$ and stays that way throughout, the one for $p$-LFC stays at zero throughout. The power of the tests based on $p$-LFC drops to zero when the critical value equals the sample size (number of groups in this case) and hence the CDF at the critical value equals one. When this happens, the term after the ``+" sign in $(\ref{eq:r5})$ equals zero and hence the CDF of $P^{rand1}$ at $\alpha$ rises to $\alpha$. In general, the power of tests based on the four $p$-values increases with an increase in group size when only the number of inspections is fixed and decreases with an increase in group size when only the number of individuals is fixed. These two findings are analogous to the findings of \cite{tebbs2003more} who used LRT and \cite{tebbs2006hypothesis} who used LRT, angular-transformed, and Bartholomew's statistics in testing for simply ordered proportions in group testing. 
\end{enumerate}
\end{example}



\section{ESTIMATION OF THE PROPORTION OF TRUE NULL HYPOTHESES}
\label{s:multiple}

In this section we assume that we have a population of size $N$ that can be divided into $k >1$ categories. The items within each category can be classified as positive (denoted by 1) if they contain a certain trait of interest or negative (denoted by 0) if they do not possess this trait.  The status of an item in category $i$ is therefore distributed as Bernoulli with mean $\theta_i$, where $\theta_i$ is the probability that an item is positive within the $i^{th}$ category, $i=1,\ldots,k$. 

For $i \in \{1, \ldots, k\}$, let $n_i$ be the number of items in the  $i^{th}$ category. Further, let $Q_{i\ell}=1$ if the $\ell^{th}$ item in the $i^{th}$ category is positive and $Q_{i\ell}=0$ otherwise, for $i=1,\ldots,k$ and $\ell=1,\ldots,n_i$. For each $i$, we assume that the sampling scheme is such, that $Q_{i,1}, \ldots Q_{i,n_i}$ are i.i.d. Bernoulli random variables with mean $\theta_i$. Define $R_i=\sum_{\ell=1}^{n_i} Q_{i\ell}$, which is the number of positive items in the $i^{th}$ category, then, $R_i\sim Bin (n_i,\theta_i)$. The MLE for $\theta_i$ is $\widehat{\theta}_i=r_i/n_i$, which is the observed proportion of positive items in the $i^{th}$ category. 

In this section, we consider testing problems referring to all the $k$ categories simultaneously. This leads to a multiple testing problem with $k$ tests, the source of multiplicity being the division of the population into $k$ categories. The $k$ pairs of hypotheses to be tested are $H_i: \theta_i\leq \theta^*_i$ versus $K_i: \theta_i>\theta^*_i$, where each $\theta_i^{*}$ is a pres-specified constant, $1\leq i \leq k$. 
The resulting (binomial) test statistics are denoted by $T(\pmb{Q_1}),...,T(\pmb{Q_k})$, where $\pmb{Q_i}=(Q_{i,1},...,Q_{i,n_i})^{\top}$ for the $i^{th}$ category. 

Assume that we have different sets $\{U_1,...,U_k\}$ and $\{\tilde{U_1},...,\tilde{U_k}\}$ of i.i.d. uniform variables on the interval $[0,1]$. Assume also that a set of constants $\{c_1,...,c_k\}$ with $c_i \in [0,1]$ for all $i \in \{1, \ldots, k\}$ is given. In practice, it is often preferred to choose $c_1 = c_2 = \ldots = c_k = c$ for simplicity; cf. \cite{hoang2022randomized}. 
Using the above notations, we obtain $k$ LFC $p$-values $\{P_1^{LFC}(\pmb{Q_1}),...,P_k^{LFC}(\pmb{Q_k})\}$ with the corresponding $k$ single-stage randomized $p$-values 
$\{P_1^{rand1}(\pmb{Q_1},U_1,c),...,P_k^{rand1}(\pmb{Q_k},U_k,c)\}$. Similarly, we also have $k$ UMP $p$-values $\{P_{1,T}^{rand}(\pmb{Q_1},U_1),$ $...,P_{k,T}^{rand}(\pmb{Q_k},U_k)\}$ with the corresponding $k$ two-stage randomized $p$-values 
$\{P_1^{rand2}(\pmb{Q_1},U_1,\tilde{U_1},c),$ $...,P_k^{rand2}(\pmb{Q_k},U_k,\tilde{U_k},c)\}$.

We denote the multiple test for $H_1,...,H_k$ by $\varphi=(\varphi_i:1\leq i\leq k)$ and the utilized $p$-values by $\{p_1,...,p_k\}$. By convention, $\{\varphi_i=1\}$ denotes the event that $H_i$ is rejected, while $\{\varphi_i=0\}$ denotes the event that $H_i$ is retained. 
Let $I_0\equiv I_0(\boldsymbol{\theta})\subseteq I=\{1,\ldots, k\}$ denote the index set of true null hypotheses under $\boldsymbol{\theta} = (\theta_1, \ldots, \theta_k)^\top$. Let $V(\varphi)=\sum_{i\in I_0}\varphi_i$ be the number of type I errors (false rejections). Define the family-wise error rate (FWER, cf. \cite{hochberg1987multiple}, page 3) of $\varphi$ under $\boldsymbol{\theta}$ by $\text{FWER}_{\boldsymbol{\theta}}(\varphi)=\mathbb{P}_{\boldsymbol{\theta}}(V(\varphi)>0)$. This is the probability of at least one false rejection of $\varphi$ under $\boldsymbol{\theta}$. We say that the FWER is controlled at level $\alpha$ by the multiple test $\varphi$ if $\mathrm{sup}_{\boldsymbol{\theta}\in \Theta} \text{FWER}_\theta(\varphi)\leq \alpha$, where $\Theta = [0, 1]^k$ is the parameter space pertaining to $\boldsymbol{\theta}$. 

One multiple test controlling the FWER (without further conditions) is given by the widely used Bonferroni correction, meaning that each individual test $\varphi_i$ is carried out at the multiplicity-adjusted (local) level $\alpha/k$, for $i \in \{1, \ldots, k\}$. If all the $k$ marginal test statistics are jointly independent (as it is the case in our model), the {\v{S}}id{\'a}k correction can be used, meaning that the local level is given by $1-(1-\alpha)^{1/k}$, which is slightly larger than $\alpha / k$. In both cases,  $\varphi_i = 1$ if and only if $p_i \leq \alpha_{adj}$, for all $i\in I$, where $\alpha_{adj}$ equals $\alpha/k$ or $1-(1-\alpha)^{1/k}$ for the Bonferroni or the {\v{S}}id{\'a}k test, respectively. 

Let $k_0 = k_0(\boldsymbol{\theta}) = |I_0(\boldsymbol{\theta})| \leq k$ denote the number of true null hypotheses (under $\boldsymbol{\theta}$). Knowledge of this quantity is in itself of scientific relevance, but can also be used for enhancing the power of the Bonferroni or the {\v{S}}id{\'a}k test, respectively. Namely, replacing $k$ by $k_0$ in the definition of $\alpha_{adj}$ still leads to FWER control under the respective assumptions. However, the value of $k_0$ depends on the value of $\boldsymbol{\theta}$ and thus, $k_0$ is often unknown in practice. Therefore, it has been proposed in previous literature to utilize a pre-estimate $\hat{k}_0$ instead of $k_0$. In the case that $\hat{k}_0<k$, the power of the aforementioned multiple tests is increased when replacing $k$ by $\hat{k}_0$ definition of $\alpha_{adj}$. This methodology has been called Bonferroni plug-in (BPI) by \cite{finner2009controlling}, and the authors proved that the BPI procedure works well for independent test statistics like the ones we are considering. 


One classical, but still commonly used estimator for $k_0$ is the \cite{schweder1982plots} estimator. It is given by 
\begin{equation}\label{schweder-def}
\hat{k}_0\equiv \hat{k}_0(\lambda)= k \cdot  \frac{1-\hat{F}_k(\lambda)}{1-\lambda},
\end{equation}
where $\lambda\in [0,1)$ is a tuning parameter and $\hat{F}_k$ is the empirical CDF (ecdf) of the $k$ marginal $p$-values.  
One crucial prerequisite for the applicability of this estimator is that the marginal $p$-values $p_1, \ldots, p_k$ are (approximately) uniformly distributed on $[0,1]$ under the null hypothesis; see, e.\ g., \cite{dickhaus2013randomized}, \cite{hoang2021usage} and the references therein for details.
The randomized $p$-values considered in this work are close to meeting the aforementioned uniformity assumption, whereas the non-randomized $p$-values computed under LFCs are over-conservative when testing composite null hypothesis, especially in discrete models. Typically, the estimated value of $k_0$ becomes too large if many null $p$-values are conservative and the estimator from \eqref{schweder-def} is employed.  

In the remainder of this section, we use the publicly available Coronavirus Disease 2019 (COVID-19) data taken from  \url{https://github.com/CSSEGISandData/COVID-19} (cf. \cite{dong2020interactive}). The dataset contains confirmed COVID-19 cases and deaths as of first January 2021 for the United States of America. The data have been stratified by different regions. We omit the stratum corresponding to the American Samoa region since it has zero confirmed cases and hence zero deaths.  This implies that $k=57$ regions are used for our data analyses. 
We compare the death rates across the strata to a common death rate which we take to be our $\theta^*$ (see below).

We have performed Monte Carlo simulations to assess the (average) performance of the randomized $p$-values $P_T^{rand}$, $P^{rand1}$, and $P^{rand2}$ in the estimation of $k_0$ in this context. The tuning parameter $\lambda$ appearing in \eqref{schweder-def} has been set to $1/2$, and the constant $c$ has also been set to $1/2$ in all simulations. Furthermore, we have set $\theta_1^* =\theta_2^* = \ldots = \theta_k^* = \theta^*$. 
The (empirical) proportions of positive individuals from the data set are assumed to be the true proportions, such that the ground truth is (assumed to be) known in the simulations. However, this information is of course unknown to the practitioner. For different values of $\theta^*$ we obtain different values of $k_0$. Namely, $k_0 \in \{0, \ldots, 57\}$ increases with increasing $\theta^*$. For exemplary purposes, we present four different choices of $\theta^*$ in Table $\ref{tabletwo}$. In each of the four cases, the estimated value of $k_0$ is averaged over the $10{,}000$ Monte Carlo repetitions in the cases that randomized $p$-values are utilized in \eqref{schweder-def}. In the case of using $p$-LFC, the estimator from \eqref{schweder-def} is computed only once per chosen value of $\theta^*$, because no randomization is involved in $p$-LFC.

\begin{table}[htb]
\begin{center}
\caption{Estimates of the number of true null hypotheses.}
\label{tabletwo}
\begin{tabular}{llllll}
\rowcolor{gray} 
$\theta^*$ &$k_0$ &$\hat{k}_0^{LFC}$& 
$\hat{k}_0^{rand1}$& $\hat{k}_0^{UMP}$& $\hat{k}_0^{rand2}$\\
\rowcolor{lightgray} 
0.0100&7 &13.99 & 14.91 &13.66 &12.33 \\
0.0144&26&54.00&27.24&52.18&27.75\\
\rowcolor{lightgray} 
0.0198 &45 &89.99  &47.23 & 89.51&46.93 \\
0.0254&51&102.00&53.33&101.98&53.32\\
\end{tabular}
\end{center}
\end{table}


From Table \ref{tabletwo},  when  $\theta^*=0.0100$, the number of true null hypotheses is $7$. The $p$-value, $P^{rand2}$ gives the best estimates in terms of being on average close to the true value $k_0$. In this case, the single stage randomized p-value $P^{rand1}$ gives the worst estimate. The other two $p$-values are competing. For relatively high values of  $\theta^*$, for example,  $\theta^*=0.0254$,  $P^{rand2}$ still gives the best estimate followed by $P^{rand1}$. The $p$-values $p$-LFC and $P_T^{rand}$ lead to the worst estimates of $k_0$. This phenomenon continues throughout as the value of $\theta^*$ is increased.  

\section{DISCUSSION}
\label{s:discussion}

We have investigated how to use single- and two-stage randomized $p$-values in the context of hypothesis testing problems involving composite null hypotheses under discrete models. We have also given an example of an application of the $p$-values in group testing and in multiple testing. In single-stage randomization, the discreteness of the $p$-value is partially removed. In two-stage randomization, the discreteness of the $p$-value is removed in the first stage. This is evidenced by the CDF for $P^{rand2}$ being a smooth curve unlike the one for $P^{rand1}$ which has minor steps. 
In the second stage, another randomization has been applied to deal with the conservativity of the $p$-value $P_T^{rand}$ that arises due to the composite nature of the null hypothesis. 

The two-stage randomized $p$-value $P^{rand2}$ is the least conservative one, almost exhausting the  significance level of the tests under the null hypothesis. The non-randomized $p$-LFC is the most conservative one, but tests based on it are more powerful than those based on $P^{rand1}$. A comparison between power of the tests based on the four $p$-values and different sample sizes has also been carried out. The power function of tests based on $P^{rand2}$ is a smooth curve that increases monotonically with the sample size, while the power of tests based on $p$-LFC is not monotonically increasing in the sample size. This paradox of the power function of the tests based on $p$-LFC was described in detail by \cite{finner2001ump}. The behavior of the power function of tests based on $P^{rand2}$ can facilitate sample size planning since we are certain that including additional observational units into the study can never result in a decrease of the power of the test.

The power of the tests based on $P^{rand2}$ increases in $c$. Hence, $P^{rand2}$ is stochastically ordered in $c$, while the single-stage randomized $p$-value $P^{rand1}$ is not stochastically ordered in $c$. This could be due to the fact that $P^{rand2}$ is based on a continuous $p$-value $P_T^{rand}$ while $P^{rand1}$ is based on a discrete $p$-value $p$-LFC. Under the null hypothesis, $P^{rand2}$ becomes less conservative as $c$ decreases. It has also been demonstrated graphically that for single-stage randomized $p$-value $P^{rand1}$, there is an advantage in using a $c$ in the support of the LFC-based $p$-value $p$-LFC.  

We have given an example showing how the four $p$-values can be applied in the context of group testing when the proportion of positive individuals in the population is of interest and the inspection scheme is not perfect. When the number of inspections is fixed and the total number of individuals varies, inspecting items in pools increases the power of all the tests based on the four $p$-values. This power further rises with using larger group sizes and only drops when the group sizes are too large or the proportion of positive individuals in the population is too high. These findings are in line with previous literature. 
For relatively high prevalence, group testing should be used only with a smaller group size. For high prevalence, group testing should not be considered but instead one should revert to individual inspection. This is because of the massive drop in power of the tests based on the four $p$-values, which implies that we are less likely to make the right decisions using group testing at a high prevalence. For a fixed number of inspections, testing items in groups is recommended especially when using tests based on $P^{rand1}$ and $P^{rand2}$. This is because testing items in groups will increase the power of the tests based on these two $p$-values.
This example also considers a composite null hypothesis in group testing when the total number of individuals is fixed but the number of groups (inspections) is not fixed. The power of the tests based on the four $p$-values are slightly higher when using individual inspection than when using group testing. The slight gain in power compared to the additional costs when using individual inspection still makes group testing remain preferred. 

The simultaneous testing of  composite null hypotheses for a stratified data set with FWER control has also been considered. We have used a COVID-19 data set to illustrate the performance of the four $p$-values in the estimation of the number of true null hypotheses. In general and especially for large $\theta^*$, $P^{rand2}$ performs well compared to the other three $p$-values.

In this research we have considered upper-tailed tests. For lower-tailed tests only the definition of $p$-LFC has to modified, and the computations follow the same steps. Namely, for lower-tailed tests $p$-LFC is defined as $P^{LFC}(\pmb{X})=F_{\theta^*}(T(\pmb{X}))$. 
In future research, it would be interesting to extend our procedures to the case of positively dependent $p$-values as was done in \cite{dickhaus2012analyze}.

\vspace*{1pc}

\noindent {\bf{ACKNOWLEDGEMENT}}

\noindent{Financial support by the German Research Foundation (DFG) via Grant No. DI 1723/5-1 is gratefully acknowledged.}

\vspace*{1pc}

\noindent {\bf{CONFLICT OF INTEREST}}

\noindent {The authors have declared no conflict of interest.}

\vspace*{1pc}

\noindent{\bf{ORCID}}

\noindent{Daniel Ochieng 
\href{https://orcid.org/0000-0002-1023-5028}{\textcolor{blue}{https://orcid.org/0000-0002-1023-5028}}}

\noindent{Anh-Tuan Hoang 
\href{https://orcid.org/0000-0002-0979-4914}{\textcolor{blue}{https://orcid.org/0000-0002-0979-4914}}}

\noindent{Thorsten Dickhaus 
\href{https://orcid.org/0000-0003-3084-3036}{\textcolor{blue}{https://orcid.org/0000-0003-3084-3036}}}




\vspace*{3pc}
\renewcommand\refname{\textbf{REFERENCES}}
\makeatletter
\@temptokena{}
\makeatother

\phantom{aaaa}
\end{document}